\documentclass{PoS}
\usepackage{amssymb}
\usepackage{fontenc}
\usepackage{times}
\usepackage{mathptmx}
\usepackage{graphicx}
\usepackage{subfigure}

\title{Bulk transitions of twelve flavor QCD and $U_A(1)$ symmetry}

\ShortTitle{Bulk transitions of twelve flavor QCD and $U_A(1)$ symmetry}

\author{Albert Deuzeman\\
        Albert Einstein Center for Fundamental Physics - University of Bern, Switzerland\\
        E-mail: \email{deuzeman@itp.unibe.ch}}

\author{Maria Paola Lombardo\\
        INFN - Laboratori Nazionali di Frascati, I-00044, Frascati (RM), Italy\\
        E-mail: \email{mariapaola.lombardo@lnf.infn.it}}

\author{\speaker{Tiago Nunes da Silva} and Elisabetta Pallante\\
       Centre for Theoretical Physics - University of Groningen, The Netherlands\\
       E-mail: \email{t.j.nunes@rug.nl} and \email{e.pallante@rug.nl}}

\abstract{We present an update on our ongoing study on the nature of the bulk transition observed at strong coupling in the SU(3) gauge theory with $N_f = 12$ flavors in the fundamental representation. We show evidence that there is a first order chiral symmetry breaking bulk transition separating a region at weak coupling where chiral symmetry is restored from a region at strong coupling where chiral symmetry is broken. We also discuss hints of a separate partial restoration of $U_A(1)$ at weaker coupling. The results are in agreement with restoration of conformality in non abelian gauge theories as the number of flavors is increased.} 

\FullConference{XXIX International Symposium on Lattice Field Theory \\
                July, 10 - 16 2011\\
                Squaw Valley, Lake Tahoe, California}

\begin{document}

\maketitle

\begin{abstract}

\end{abstract}

\section{Introduction}

  In recent years attention has been drawn to studies of conformal symmetry restoration in non abelian gauge theories. On the one hand, there is theoretical interest in uncovering their phase diagram. On the other hand, the start of LHC activities creates the possibility of studying strongly coupled Physics beyond the Standard Model at the TeV scale. In particular, some models that allow for electroweak symmetry breaking live in a quasi-conformal region of the parameter space. This is the case, for example, for walking technicolor. 
A first scenario for conformal symmetry restoration in SU(N) gauge theories, for which the theory develops a non-trivial infra-red fixed point (IRFP), was conjectured by Banks and Zaks \cite{Banks:1981nn}. Later on, the conformal window scenario was introduced in \cite{Miransky:1996pd, Appelquist:1996dq}, where a conformal phase transition occurs at $N_f = N_f^c < N_f^A$, with $N_f^A$ the point where asymptotic freedom is lost. Inside this window $N_f^c < N_f < N_f^A$ chiral symmetry is exact and theories are deconfined. Because for $N_f \rightarrow N_f^c$ the IRFP occurs at increasingly stronger coupling, its study must rely on non-perturbative techniques. 

Based on \cite{Miransky:1996pd}, a strategy for a lattice study was implemented in \cite{Deuzeman:2009mh}, favoring the existence of a conformal window opening at $N_f^c \lesssim 12$. The study also observed a lattice bulk transition to a chirally broken phase at strong bare lattice coupling. A study of the nature of this transition is required for a better understanding of the scenario. In the first place, it is important to distinguish between a bulk (zero temperature) and a thermal (finite non-zero temperature) chiral transition. The latter only occurs for theories below the conformal window, signalled by a shift in the critical coupling when changing the lattice temporal extent; it is then possible to match the scaling of this shift with the one extracted from perturbation theory. In the conformal window, the transition happens at zero temperature and it is a lattice artifact referred to as a bulk transition. No scaling with the lattice temporal extent (i.e. no thermal behavior) should be observed in this case.

Another point of interest is the order of the transition.  A second order nature of the bulk transition would signal the emergence in the continuum of a second non trivial fixed point which is an attractor in the ultra-violet, and thus of a strongly coupled conformal theory different from QCD. The following scenarios are then possible. First, the transition is thermal and the theory is below the conformal window. Second, it is a second order bulk transition, what would point to a theory in the conformal window that, in addition to an infra-red fixed point (IRFP), contains an ultra-violet one (UVFP). The final alternative is to find a first order bulk transition, what points to a theory in the conformal window without a UVFP; such a fixed point, however, should still emerge as the end-point of a line of first order chiral transitions at the lower endpoint of the conformal window.   

In the remainder we report on our ongoing study of the nature of the strong coupling transition observed in simulations of SU(3) gauge theories with twelve fundamental flavors. 

\section{Simulations}

We are currently simulating an SU(3) gauge theory with twelve flavors of staggered fermions in the fundamental representation. In order to suppress lattice artifacts, we use a tree level Symanzik improved gauge action, and Kogut-Susskind (staggered) fermions with the Naik improvement scheme, which effectively extends the Symanzik improvement to the matter content. This work is an update to the study presented in \cite{Deuzeman:2010fn}. We performed runs with cold (ordered) and hot (random) starts over an extended range of bare lattice couplings on volumes $6^3x6$ and $12^3x24$ for the bare masses am = 0.020, 0.025, 0.030. For the same range of bare lattice couplings, we have also performed runs on volumes $16^3x24$ and $24^3x24$ for am=0.025. With these new runs, we have improved our statistics for the data presented in \cite{Deuzeman:2010fn}, investigated metastabilities at different lattice volumes and added to our chiral analysis results at a new bare mass below the crossover regime.  

\subsection{The Chiral Condensate}

For small enough bare masses ($am \lesssim 0.04$) at the simulated volumes two rapid crossovers are observed in the value of the chiral condensate: a large one at stronger coupling and a smaller one at weaker coupling\cite{Deuzeman:2010fn}. 
\begin{figure}[ht]
 \centering
 \subfigure[\label{fig:massdep}]%
 {\includegraphics[width=0.49\linewidth]{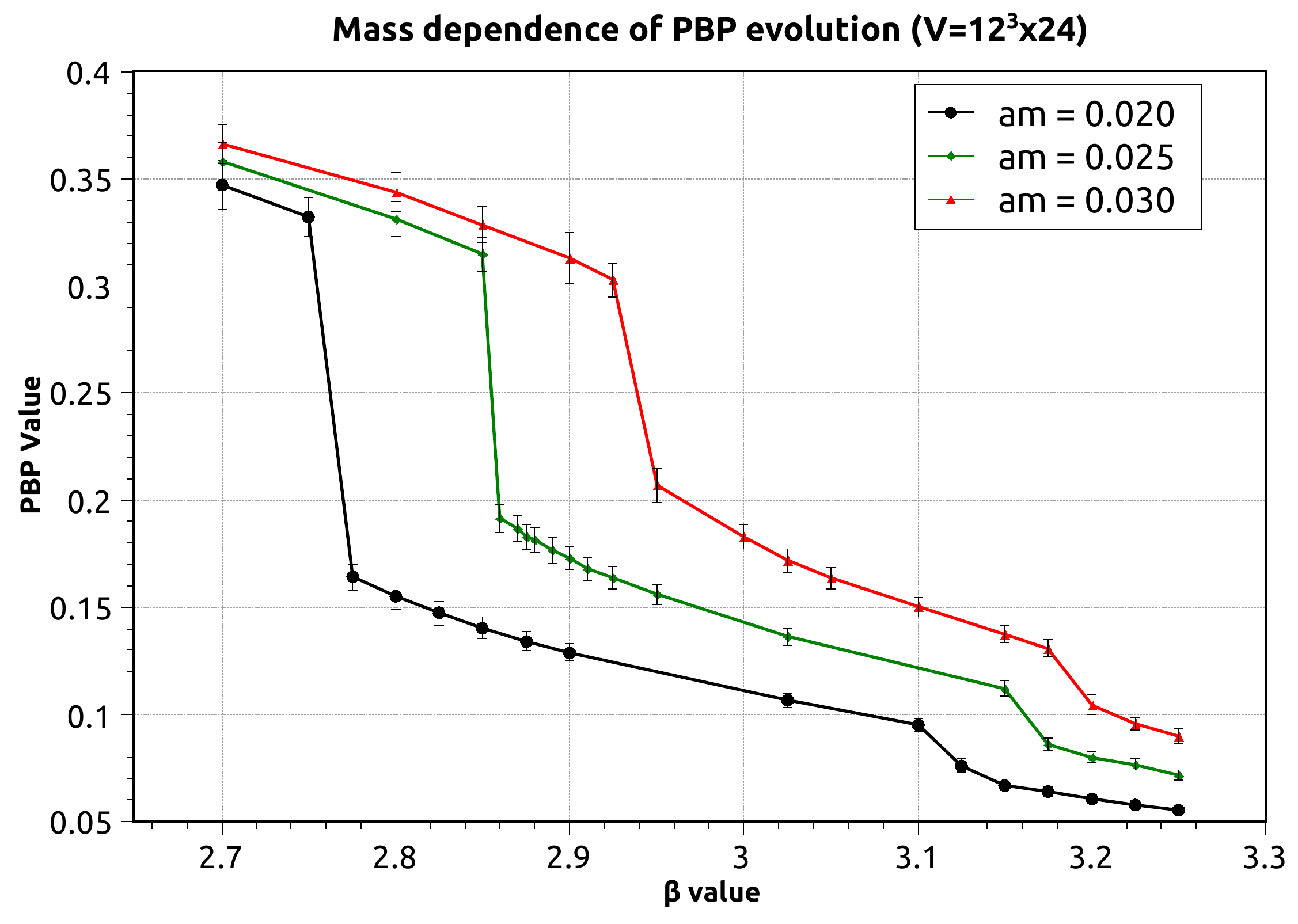}}
 \subfigure[\label{fig:ntdep}]%
 {\includegraphics[width=0.49\linewidth]{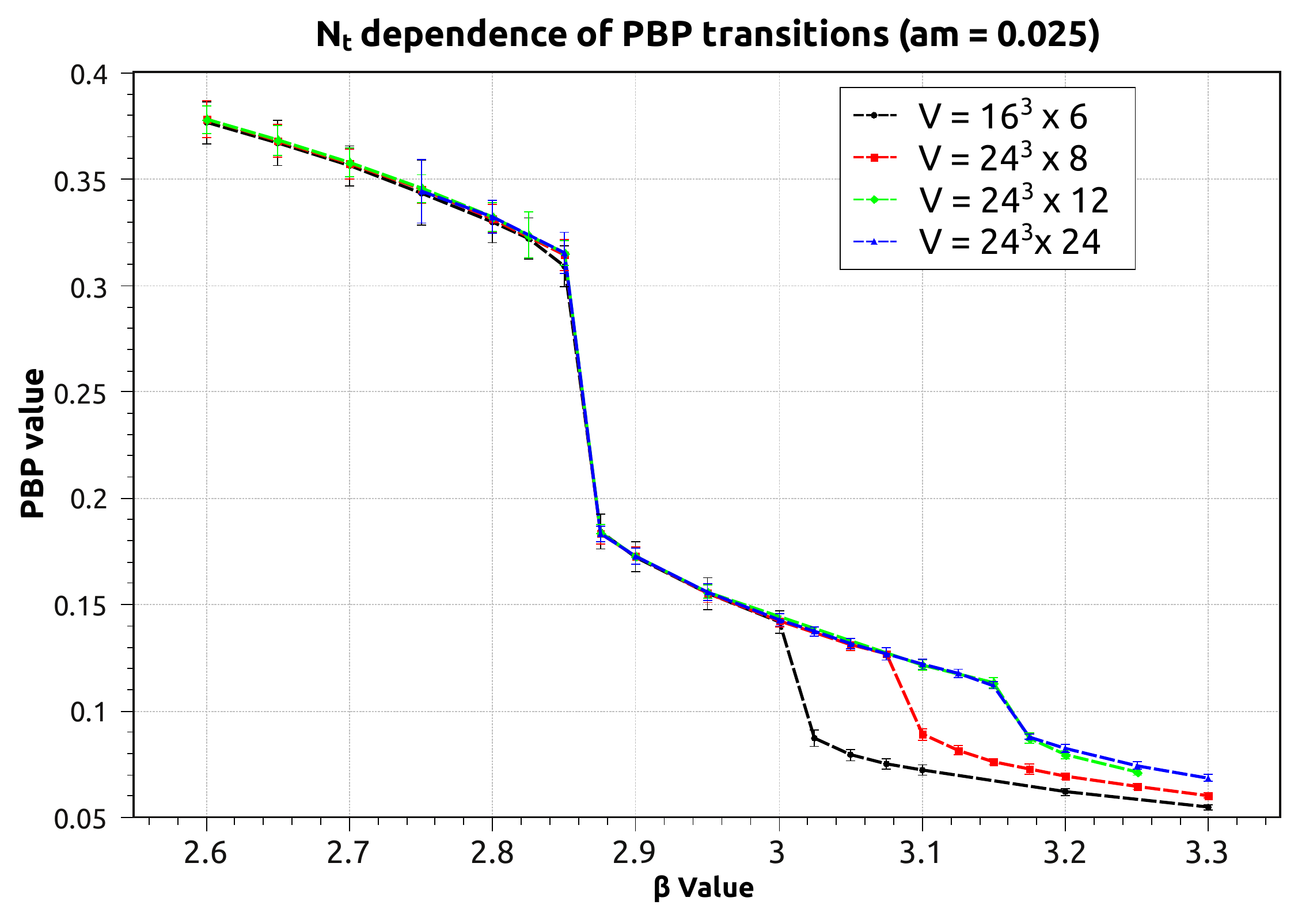}}
 \caption{(a) Rapid crossovers in the chiral condensate (PBP) as a function of the coupling $\beta = 6/g^2$  in the strong coupling region for different bare masses. (b) The behavior as the lattice temporal extent is enlarged.}
 \label{fig:first}
\end{figure}
In Figure \ref{fig:first} we show the chiral condensate for a range of couplings simulated at different bare masses and volumes. The jump at stronger coupling becomes more pronounced as the bare mass decreases. No dependence on the lattice temporal extent has been observed for this jump and no perturbative scaling can be realized. This behavior points towards a non-thermal zero-temperature (bulk)  transition.
The smaller jump at weaker coupling, differently from the strong coupling jump, tends to smooth out for lower masses. It also moves towards weaker couplings as the lattice temporal extent is enlarged up to $N_t = 10$, becoming insensitive to temporal extent variation for $N_t \gtrsim 10$ (see also \cite{Deuzeman:2010fn}). 

\begin{figure}[!ht]
  \centering
 \includegraphics[width=0.8\textwidth]{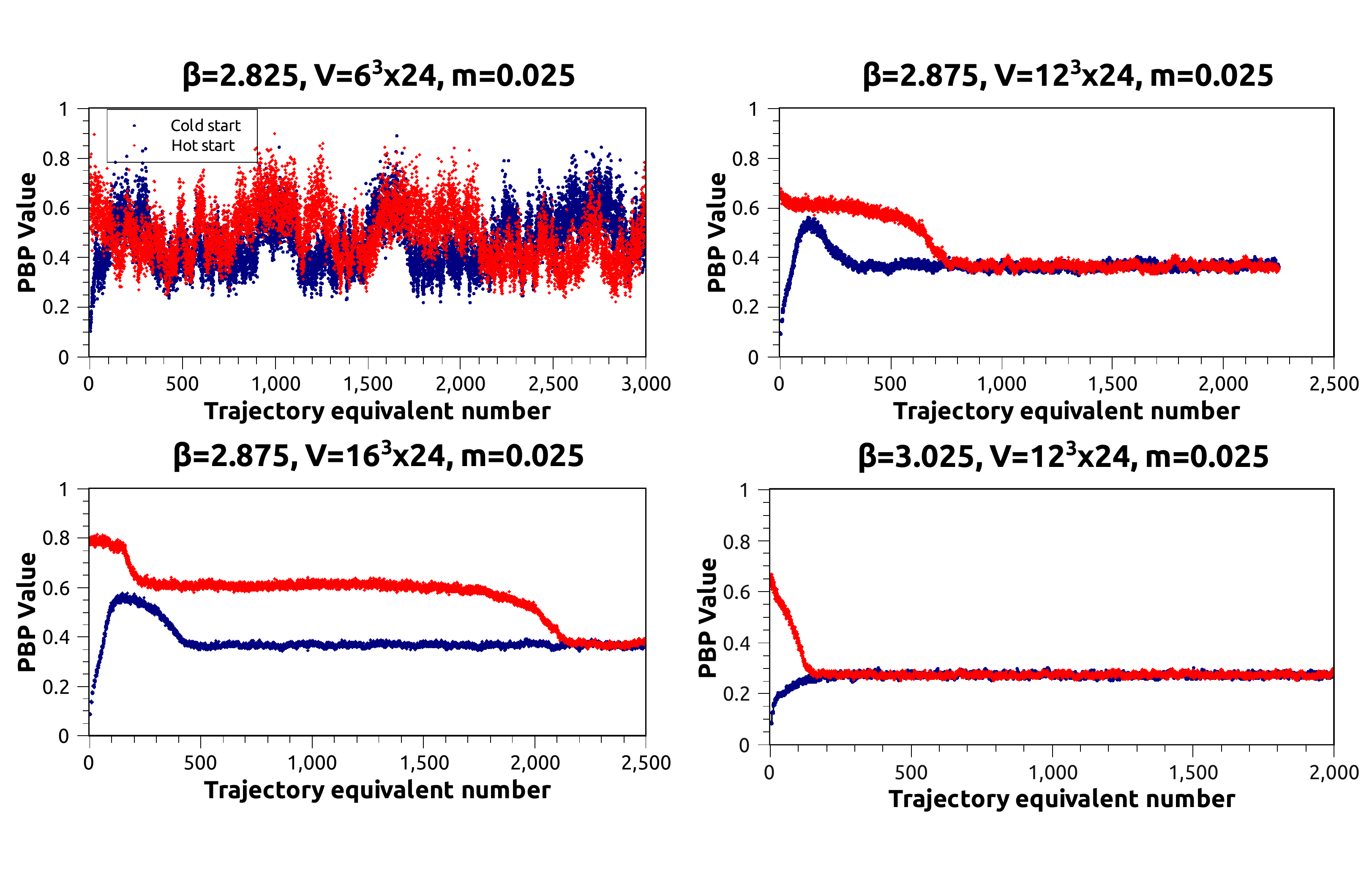}
 \caption{Comparison between cold (ordered) and hot (random) starting configurations.  As the volume is increased the stability of these metastable states is increased. Away from the transition region (including the intermediate region between the two jumps) no metastability is observed (lower-right).}
 \label{fig:tunnel24}
\end{figure}

\begin{figure}[ht]
 \centering
 \subfigure[\label{fig:hyst16x24}]%
 {\includegraphics[width=0.46\linewidth]{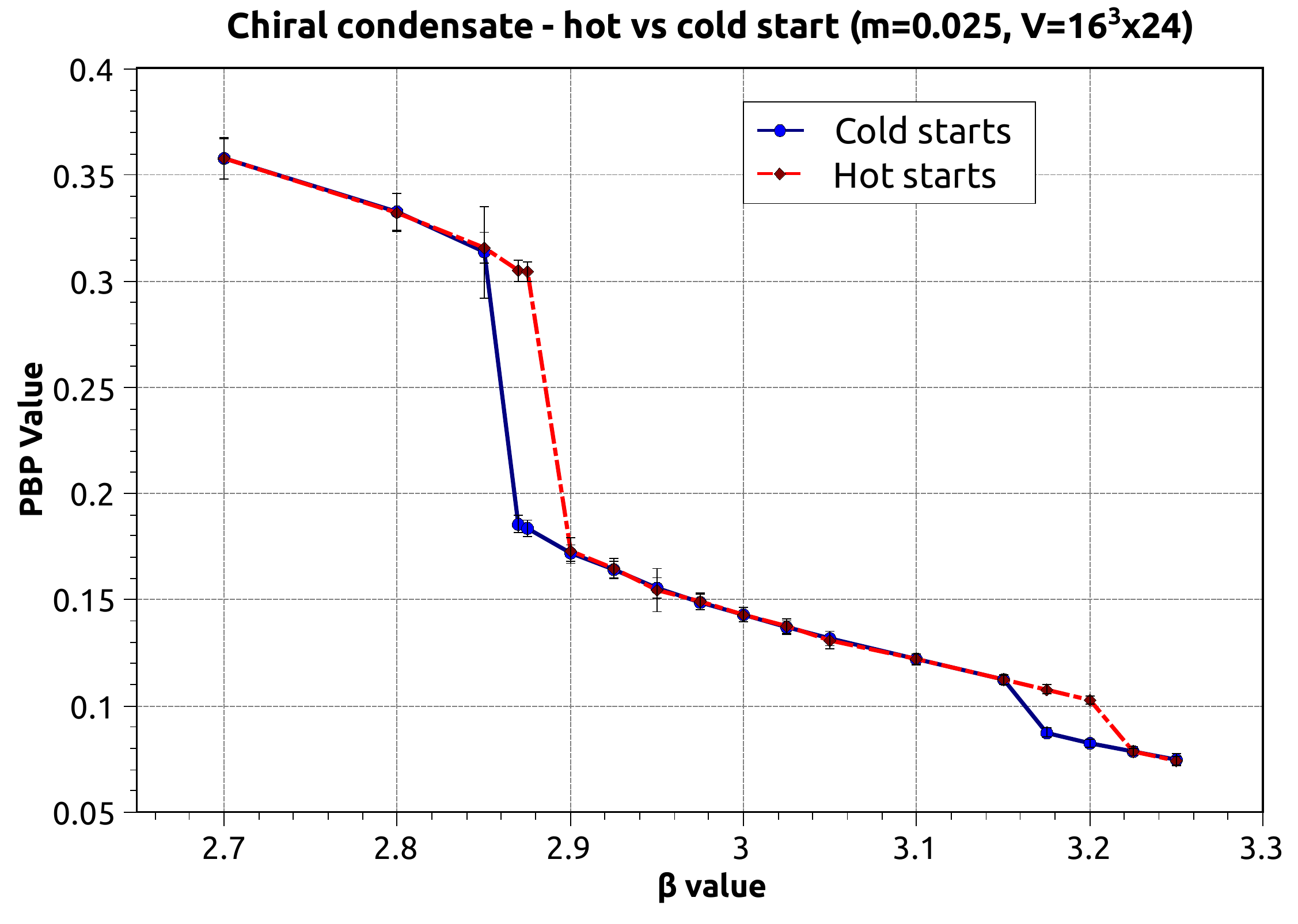}}
 \subfigure[\label{fig:massdepbc}]%
 {\includegraphics[width=0.46\linewidth]{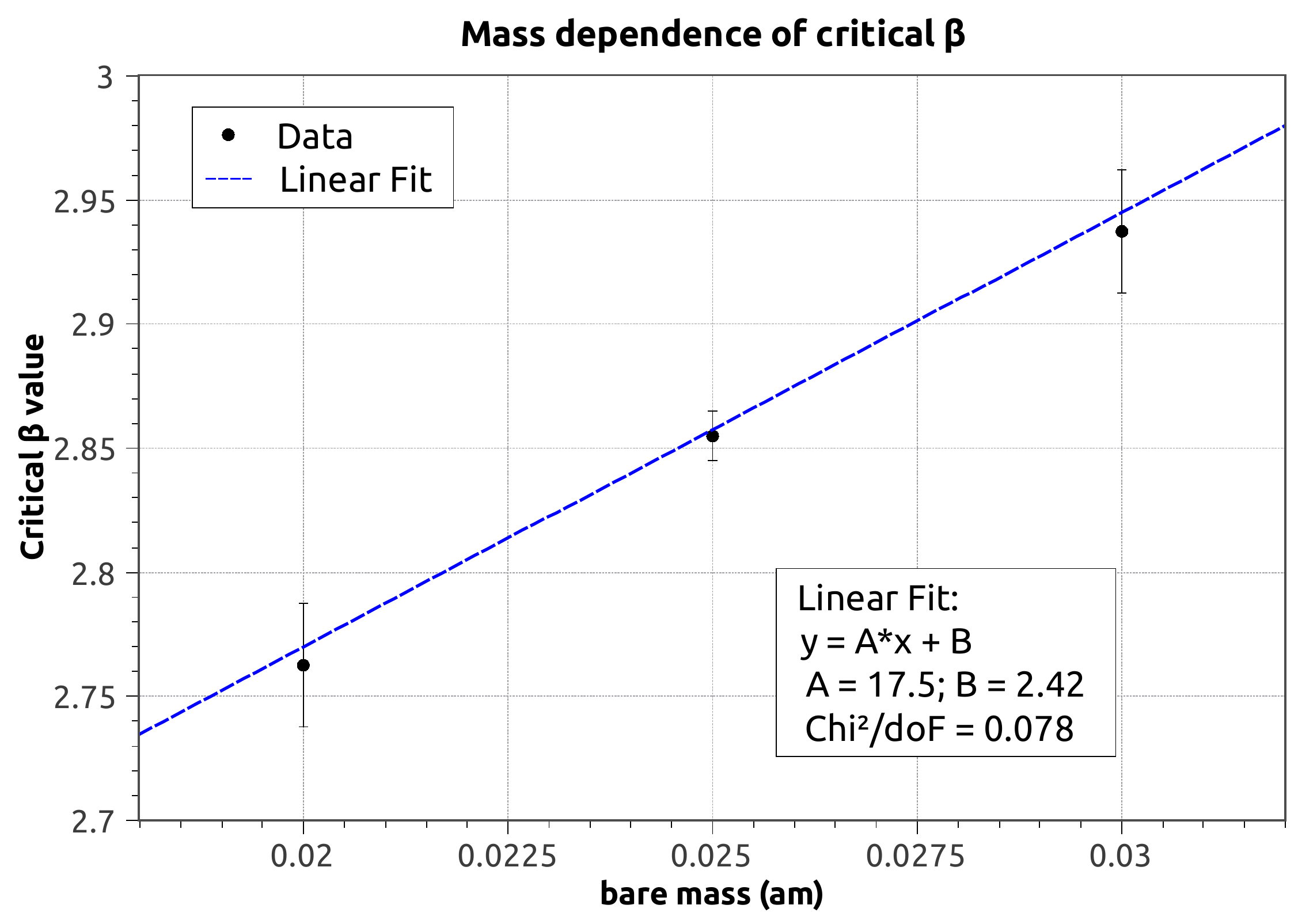}}
  \caption{(a) Hysteresis loops observed in the chiral condensate for cold (ordered) and hot (random) starting lattice configurations. (b) Mass dependence of the critical $\beta$ value extracted from the central point of the strong coupling seems in agreement with a linear scaling expected for a first order transition.}
 \label{fig:third}
\end{figure}

Figure \ref{fig:tunnel24} illustrates some representative Monte Carlo histories of the chiral condensate. Tunnelling between metastable states is observed around the transition at stronger coupling for small volumes (top-left corner). As the lattice volume is enlarged the occurrence of tunnelling ceases and the stability of metastable states is increased as we approach infinite volume (top-right and lower-left). This persistent separation between metastable states at large volumes generates the hysteresis loop shown in Figure \ref{fig:hyst16x24}. 

In \cite{Deuzeman:2010fn} we raised the hypothesis that a large hysteresis loop could exist in the intermediate region between the two jumps. However, no metastability is observed in the region between the two jumps (Figure \ref{fig:tunnel24}, lower-right). This excludes this hypothesis and indicates that the two jumps are of different nature. 

\subsection{Chiral symmetry breaking}

\begin{figure}[ht]
 \centering
 \subfigure[\label{fig:voldepc}No volume dependence is observed for the connected susceptibilities.]%
 {\includegraphics[width=0.48\linewidth]{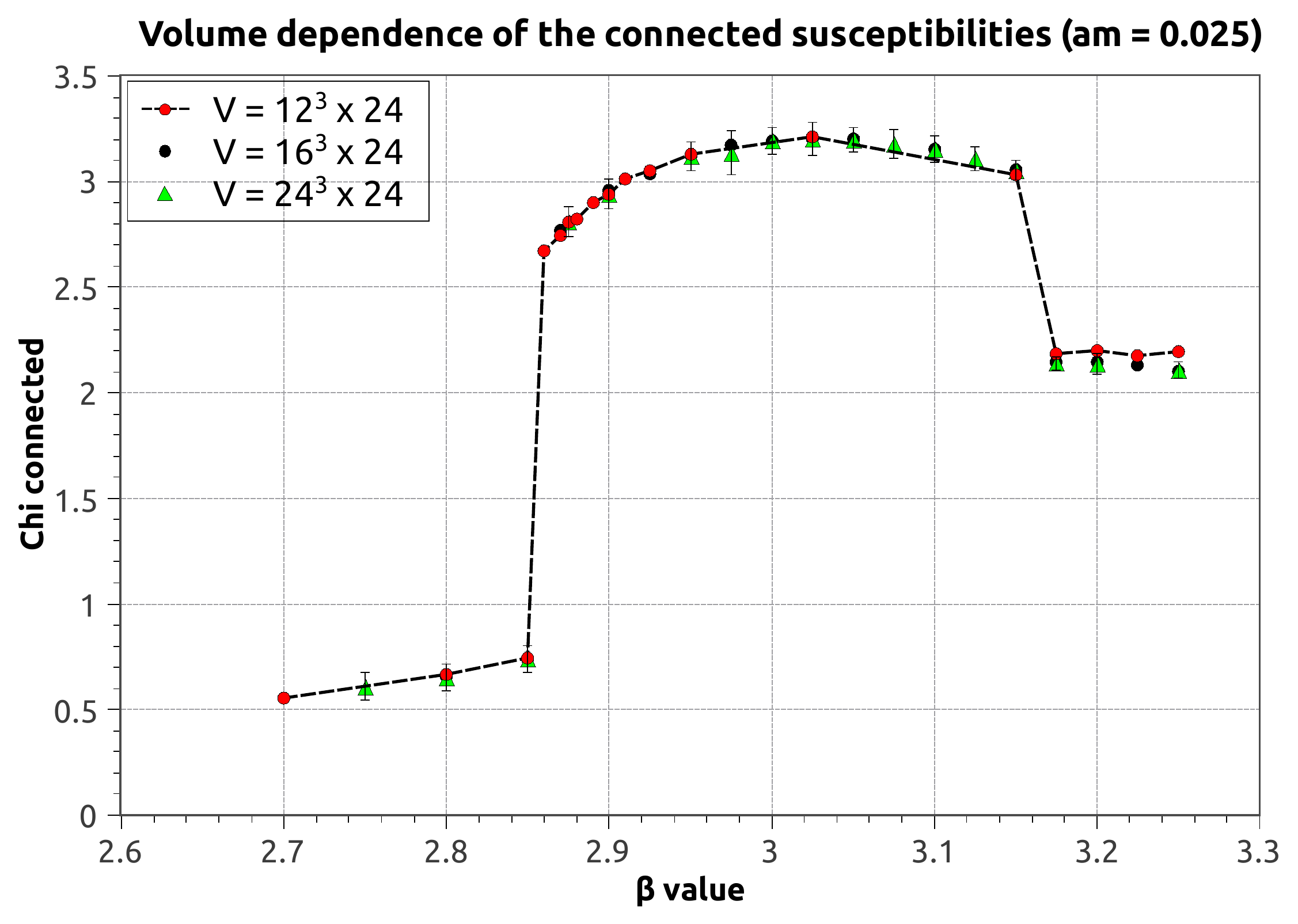}}
 \subfigure[\label{fig:voldepd}Absence of tunneling between metastable states at large volumes supresses the volume scaling of the disconnected component.]%
 {\includegraphics[width=0.48\linewidth]{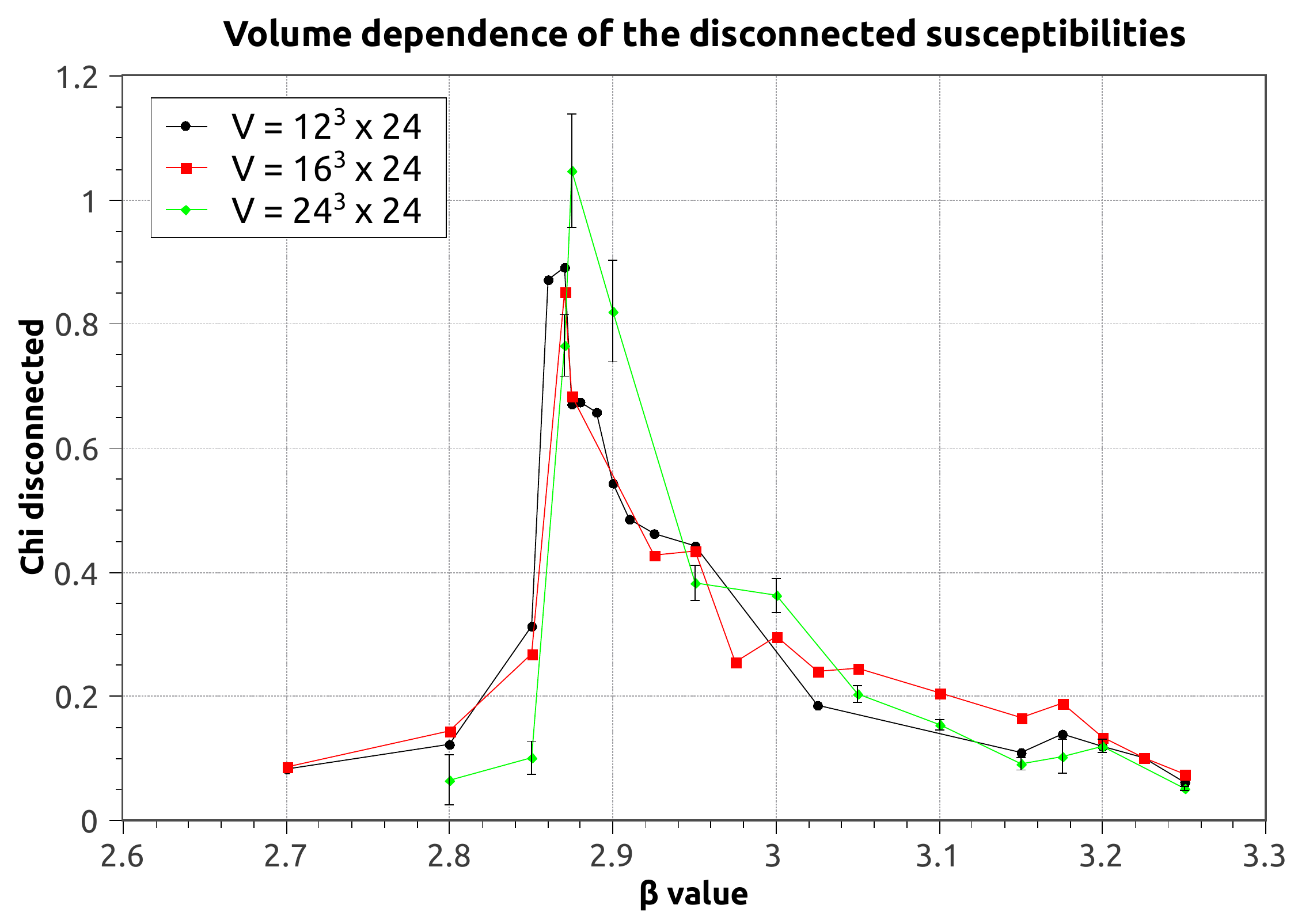}}
  \caption{Volume dependence of the connected and disconnected components of the susceptibilities. The analysis is in progress, so the results are preliminary. Error bars are shown for the largest volume.}
 \label{fig:suscvol}
\end{figure}

We now show the results for some important chiral symmetry observables, namely the connected and disconnected susceptibilities and the chiral cumulant. The connected component of the chiral susceptibility exhibits almost-discontinuities at the condensate jumps that can be possibly explained by the absence of tunneling between metastable states at the volumes at which they were extracted. They also show no volume dependence (Figure \ref{fig:voldepc}). Notice that decreasing the bare mass increases the magnitude of both discontinuities (Figure \ref{fig:third}).

\begin{figure}[ht]
 \centering
 \includegraphics[width=0.85\linewidth]{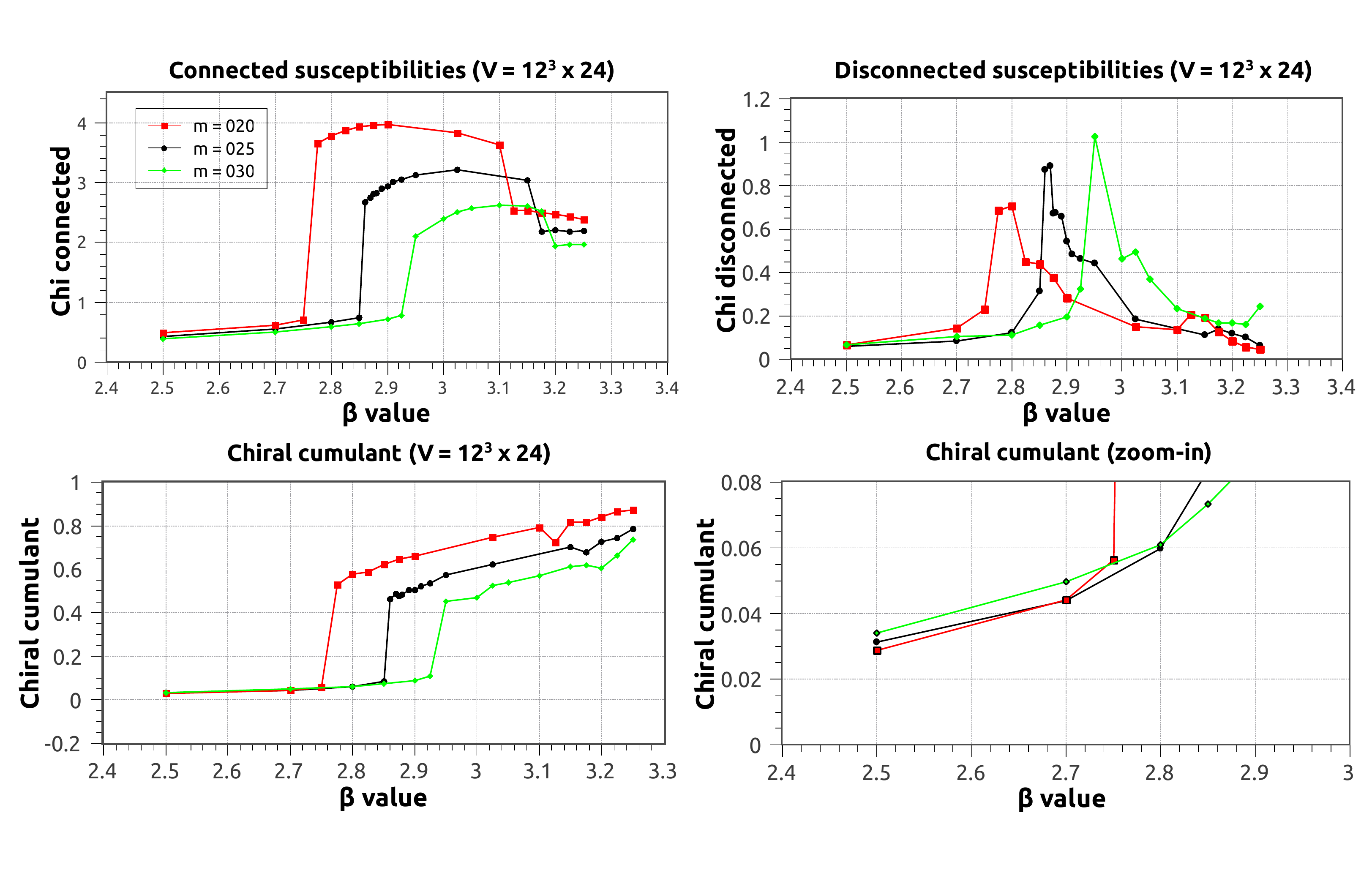}
 \caption{Mass dependence of the susceptibilities and the chiral cumulant. The cumulant goes to zero in the chirally broken region and to one in the chiral restored region. An inversion in the mass scaling of the cumulant is observed at the stronger coupling transition, as expected for a chiral symmetry breaking transition.}
 \label{fig:fourth}
\end{figure}

The disconnected component of the chiral susceptibility shows a pronounced peak in correspondence of the strong coupling jump (Figure \ref{fig:fourth}, top-right). Once the location of the peaks is acquired with a better resolution, these can be used to trace the mass dependence of the transition location. This can be roughly estimated from the central position of the jumps, as shown in Figure \ref{fig:massdepbc}. The scaling appears to be consistent with the linear behavior expected from a first order transition.

No clear volume dependence is visible for the disconnected susceptibilities from Figure \ref{fig:voldepd}. Again, this may be attributed to the absence of tunnelling between metastable states at the simulated volumes.

The chiral cumulant is defined as the ratio between the longitudinal and transverse susceptibilities: $R_{\chi}=\chi_\sigma/\chi_\pi = (\chi_{conn}+\chi_{disc})/(\langle\bar{\psi}\psi\rangle/m)$. It is expected to reach unity in the chirally restored phase and go to zero in the broken phase. Figure \ref{fig:fourth} strongly suggests that chiral symmetry breaking occurs in correspondence with the jump at strong coupling. Also, for a fixed coupling, as the bare mass is increased, the value of the cumulant should also increase in the broken phase, while it should decrease in the chirally symmetric phase. It is thus possible to locate the chiral symmetry breaking transition by looking at the point where the inversion in the mass scaling of the chiral cumulant occurs. Again, this is observed (Figure \ref{fig:fourth}, lower-right) to occur at the jump of the chiral condensate at stronger coupling. This combined evidence from the chiral cumulat strongly indicates that the jump at stronger coupling is the bulk chiral symmetry breaking transition. 

\subsection{Hints of $U_A(1)$ restoration}

Embedded in the data presented so far there is some evidence that the jump at weaker coupling is induced by restoration of $U_A(1)$ symmetry. For this transition the connected component of the chiral susceptibility, and not the chiral condensate, serves as an order parameter. Notice that while the jump in the chiral condensate smooths out and is expected to vanish with decreasing bare masses, the discontinuity in the connected susceptibilities does the opposite, enlarges with decreasing bare masses (see Figure \ref{fig:fourth}, top-left). This indicates that it will not disappear in the chiral limit. The signal of the connected susceptibility is also stable against volume variation as seen in figure \ref{fig:voldepd}. 

\begin{figure}[ht]
 \centering
 \subfigure
 {\includegraphics[width=0.48\linewidth]{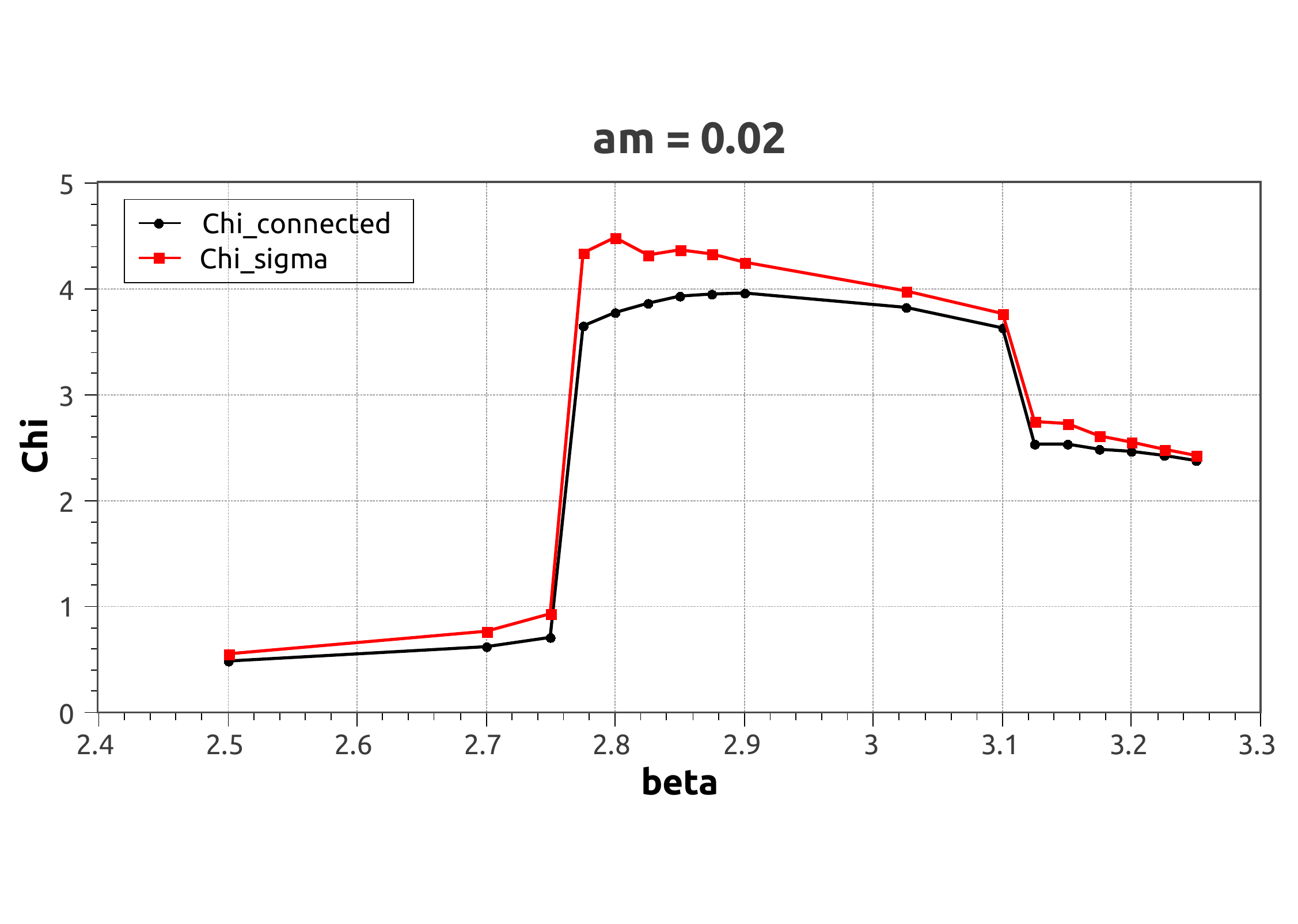}}
 \subfigure
 {\includegraphics[width=0.48\linewidth]{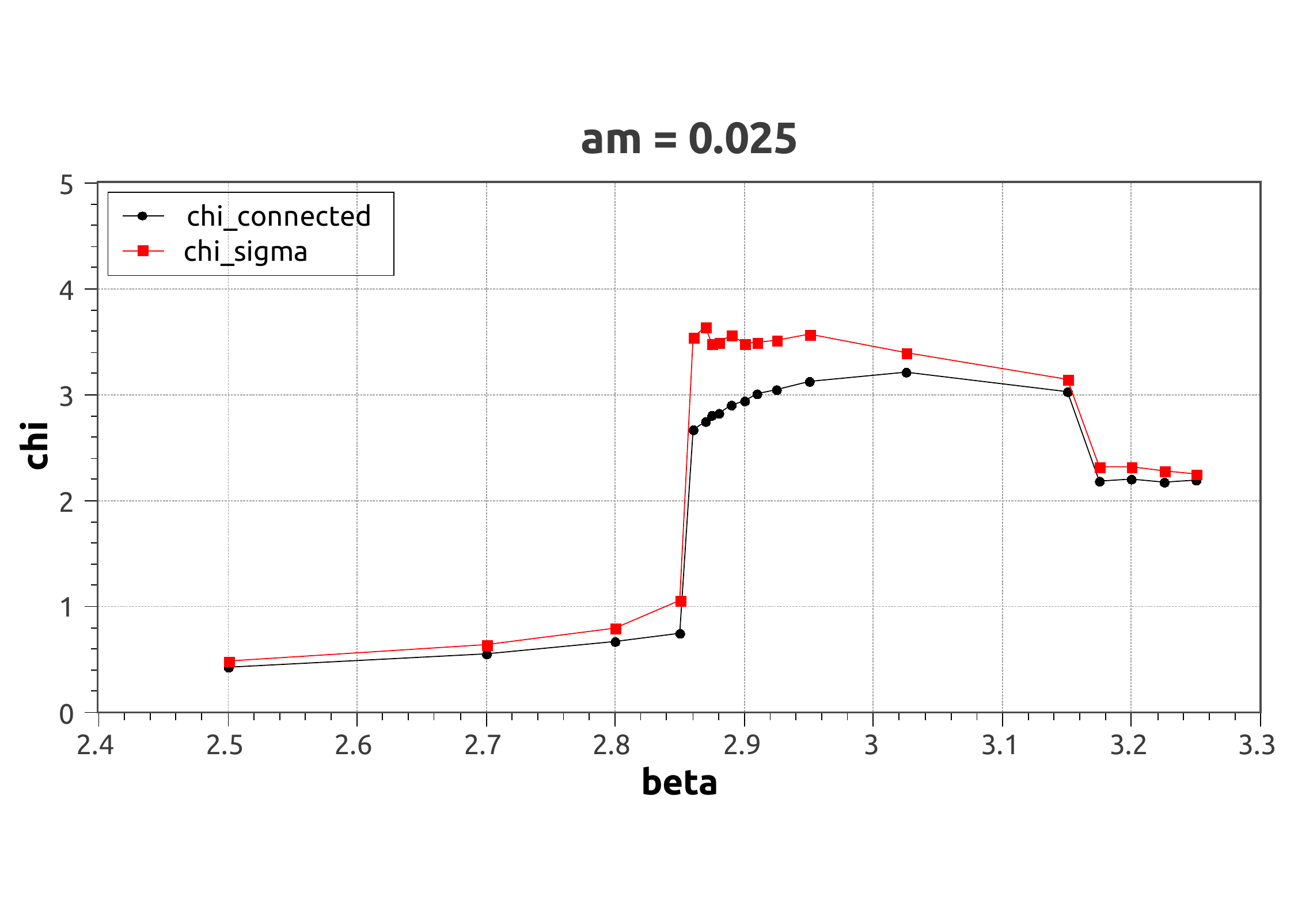}}
  \caption{Splitting between the $\delta$ and the $\sigma$ as seen from the connected and disconnected components of the susceptibilities for $am = 0.020$ and $am = 0.025$.}
 \label{fig:splitting}
\end{figure}

This behavior suggests that as one goes from strong to weak coupling $U_A(1)$ symmetry remains broken after the bulk chiral transition and gets partially 	restored\footnote{Strictly speaking, $U_A(1)$ is broken by the staggered action at non zero lattice spacing, and it can be fully restored only in the continuum limit.} afterwards. In such a scenario, one should be able to observe a splitting between the $\delta$ and the $\sigma$ in the intermediate region between the two jumps, where chiral symmetry is already restored but not $U_A(1)$. Since (modulo renormalization factors) $\chi_{\sigma} = \chi_{conn} + \chi_{disc} \propto 1/M_{\sigma}$ and $\chi_{conn} \propto 1/M_{\delta}$, it is also possible to observe this splitting from the susceptibilities as shown in Figure \ref{fig:splitting}. The splitting tends to disappear at the weakest couplings. On the strong coupling side, once chiral symmetry is broken, the $\sigma$ and the $\delta$ become heavy, the susceptibilities become small, and the splitting is suppressed. 
  
\section{Conclusion and Outlook}

We have reported on the current status of our ongoing study of the bulk transition observed for SU(3) gauge theory with twelve flavors.

The analysis of the chiral cumulant brings clear evidence that a chiral symmetry breaking transition happens at the jump at stronger coupling. For this, the absence of thermal behavior points towards a bulk transition. Also, the presence of metastability, the occurrence of tunnelling at small volumes, and increased stability of metastable states at larger volumes, the occurrence of hysteresis, and the linear mass dependence of the jump location, all point towards a first order transition.

The absence of metastability in the intermediate region excludes the possibility of a large hysteresis loop between the jumps, indicating that they are of different nature. The mass scaling of the connected component of the chiral susceptibility, the presence of a  splitting between the $\delta$ and the $\sigma$ in the region between the jumps, and its suppression at weaker coupling suggest that a partial restoration of $U_A(1)$ symmetry takes place at the jump at weaker coupling. The chiral condensate is not an order parameter for this transition, thus its jump is expected to vanish in the chiral limit since the chiral condensate in the chiral limit is zero on both sides. It can also eventually disappear for simulations at sufficiently small masses and sufficiently large volumes. 

A scenario emerges where $N_f=12$ is in the conformal window and, as one goes from stronger to weaker couplings, chiral symmetry is restored by a strong first order bulk transition and this is followed by a partial $U_A(1)$ restoration at weaker couplings. 

\section*{Acknowledgments}

This work was in part based on the MILC collaboration's public lattice gauge theory code. Computer time was provided through the Dutch National Computing Foundation (NCF) and the University of Groningen.

\end{document}